\documentclass{aastex}
\usepackage{emulateapj5}

\newcommand{\ha}{H$\alpha$}

\newcommand{\jks}{$J-K_s$}
\newcommand{\ks}{$K_s$}

\shortauthors{HALL}
\shorttitle{L DWARF WITH H-ALPHA} 
\slugcomment{Accepted to ApJL December 6, 2001}

\begin{document}

\title{2MASSI J1315309$-$264951: 
An L Dwarf with Strong and Variable H$\alpha$ Emission}

\author{
Patrick B. Hall\altaffilmark{1}
\altaffiltext{1}{
Pontificia Universidad Cat\'{o}lica de Chile, Departamento de Astronom\'{\i}a
y Astrof\'{\i}sica, Facultad de F\'{\i}sica, Casilla 306, Santiago 22, Chile,
and Princeton University Observatory, Princeton, NJ 08544-1001;
E-mail: phall@astro.puc.cl}
}

\begin{abstract}

2MASSI J1315309$-$264951 is 
an L3 dwarf with strong H$\alpha$ emission discovered in the
course of a color-selected survey for active galactic nuclei using 
the Two-Micron All-Sky Survey (2MASS).  The strength of its 
H$\alpha$ emission 
decreased by about a factor of two between two epochs separated by 137 days.
This is the first time that variable \ha\ emission has been reported in an
L dwarf, and is probably the first observation 
of an \ha\ flare in an L dwarf.
The value of $\log(L_{\rm H\alpha}/L_{\rm bol}) > -4.17$
observed at the discovery epoch is larger than that of any other L dwarf 
but comparable to that of 
2MASSI J1237392+652615, the only reported T dwarf with \ha\ emission.
The observed variability indicates that the \ha\ emission of 2MASSI
J1315309$-$264951 is powered either by magnetic fields or by accretion in a
binary system.  Spectroscopic or narrow-band \ha\ monitoring of L and T dwarfs
on timescales of hours to days would be the most useful step toward a better
understanding of their \ha\ emission mechanism(s). 

\end{abstract}


\section{Introduction}  \label{INTRO}

For many years the coolest, lowest-mass stars known were M dwarfs, but in
recent years the L and T dwarfs have extended the stellar sequence to even lower
temperatures and masses \markcite{kir99}({Kirkpatrick} {et~al.} 1999).  These three classes of low-mass stars 
and brown dwarfs exhibit distinctly different spectral features due to the 
decrease in photospheric temperature from M through L to T.
In addition, M dwarfs often show H$\alpha$ in emission. 
Down to the early M dwarfs, \ha\ activity correlates with rotation and thus
decreases with age as stars lose angular momentum over time via stellar winds.
Beyond objects of spectral type M8, however, it appears that \ha\ activity is
stronger in more massive objects, even if they are older \markcite{giz00}({Gizis} {et~al.} 2000).
The frequency of \ha\ emission peaks around type M7 and declines for
later-type L and T dwarfs, reaching zero at L5 \markcite{giz00}({Gizis} {et~al.} 2000).
However, contrary to this trend, \markcite{bur00}{Burgasser} {et~al.} (2000) 
reported the discovery of a T dwarf with strong H$\alpha$ emission.
Here I report the discovery of an L3 dwarf with similarly strong
H$\alpha$ emission.

\section{Spectroscopy}  \label{SPEC}


2MASSI\,J1315309$-$264951 (hereafter 
2MASS\,1315$-$2649) was targeted for spectroscopy as part of a color-selected
survey of AGN candidates from the Two-Micron All-Sky Survey
Second Incremental Data Relase (2MASSI).  
That 
survey will appear 
elsewhere (Hall 2002, in preparation).

Potential AGN candidates were selected from the 2MASSI Point Source Catalog
by requiring 
$\alpha < 5^h$ or $\alpha > 15^h$, $-55\arcdeg < \delta < +5\arcdeg$, 
$|b|>25\arcdeg$,
and that they not 
be affected by blending,
saturation or confusion with other objects, known artifacts or minor planets.
%
Actual AGN candidates were then selected from the list of potential candidates
using color criteria 
based on those of \markcite{bh01err}{Barkhouse} \& {Hall} (2001).
2MASS\,1315$-$2649 was included in a subsample of 
objects with \jks$>$1.5 and \ks$<$13.5.

Spectra of 2MASS\,1315$-$2649 were secured during two runs with CSPEC at
the CTIO 1.5m telescope.  Both runs used a 300 lines/mm grating blazed at
6750\,\AA, for a resolution of 8.6\,\AA\ given the 1\farcs5 slit.
Wavelength coverage was 5780$-$9270\,\AA\ at 2.89\,\AA/pixel.
Two 12-minute exposures were begun at UT 06:09 on 2001 March 30.
A 12-minute and a 24-minute exposure were obtained through cirrus
beginning at UT 23:39 on 2001 August 15.
The two-dimensional spectra were debiased and flatfielded in the usual manner
using IRAF.\footnote{The Image Reduction and Analysis Facility (IRAF) is
distributed by NOAO, 
which is operated by AURA, Inc.,
under contract to the NSF.} 
Each pair of reduced two-dimensional spectra was then coadded with
exposure time weighting and 5$\sigma$ cosmic ray rejection.

The two-dimensional spectra show a close pair of objects.
2MASS\,1315$-$2649 at $\alpha$=13:15:30.94, $\delta$=$-$26:49:51.3 (J2000) is
an optically faint L dwarf with H$\alpha$ emission, and
USNO\,J131531.23$-$264953.0 an optically brighter star slightly south of east.
Both objects were visible on the guide camera, and the east-west slit was
positioned as best as possible to obtain spectroscopy of both.  
However, unlike USNO\,J131531.23$-$264953.0, 2MASS\,1315$-$2649 is faint enough
that it may have drifted off the slit without being immediately noticed.  
The east-west separation between the objects in the two-dimensional spectra is
5\farcs5$\pm$0\farcs3, as determined from the centroids of the spatial profiles
of the two objects in the spectra from the two independent epochs.
This is significantly larger than the 3\farcs9 east-west separation expected
for the 2MASS-USNO separation of 4\farcs23 at PA 114.5\arcdeg\ E of N.
This suggests that one or both of the objects exhibited 
proper motion between the 
2MASS epoch of 1998.411 and the spectroscopic observations.
However, better data are needed to verify this 
since CSPEC delivers a 2-3 pixel FWHM spectrum
at a spatial scale of 1\farcs3/pixel.

The spectra were optimally extracted using IRAF,
with the 
USNO star's trace used for both objects.
Flux calibration was provided by 
GD108 \markcite{oke90}({Oke} 1990) in March and LTT1788 \markcite{ham94}({Hamuy} {et~al.} 1994) in August.
The spectral type of USNO\,J131531.23$-$264953.0 must be no later than K5,
since its spectrum lacks strong features.
The spectra of 2MASS\,1315$-$2649 are discussed next. 

\section{Analysis}  \label{ANAL}

\begin{figure}
\epsscale{1.80}
\plottwo{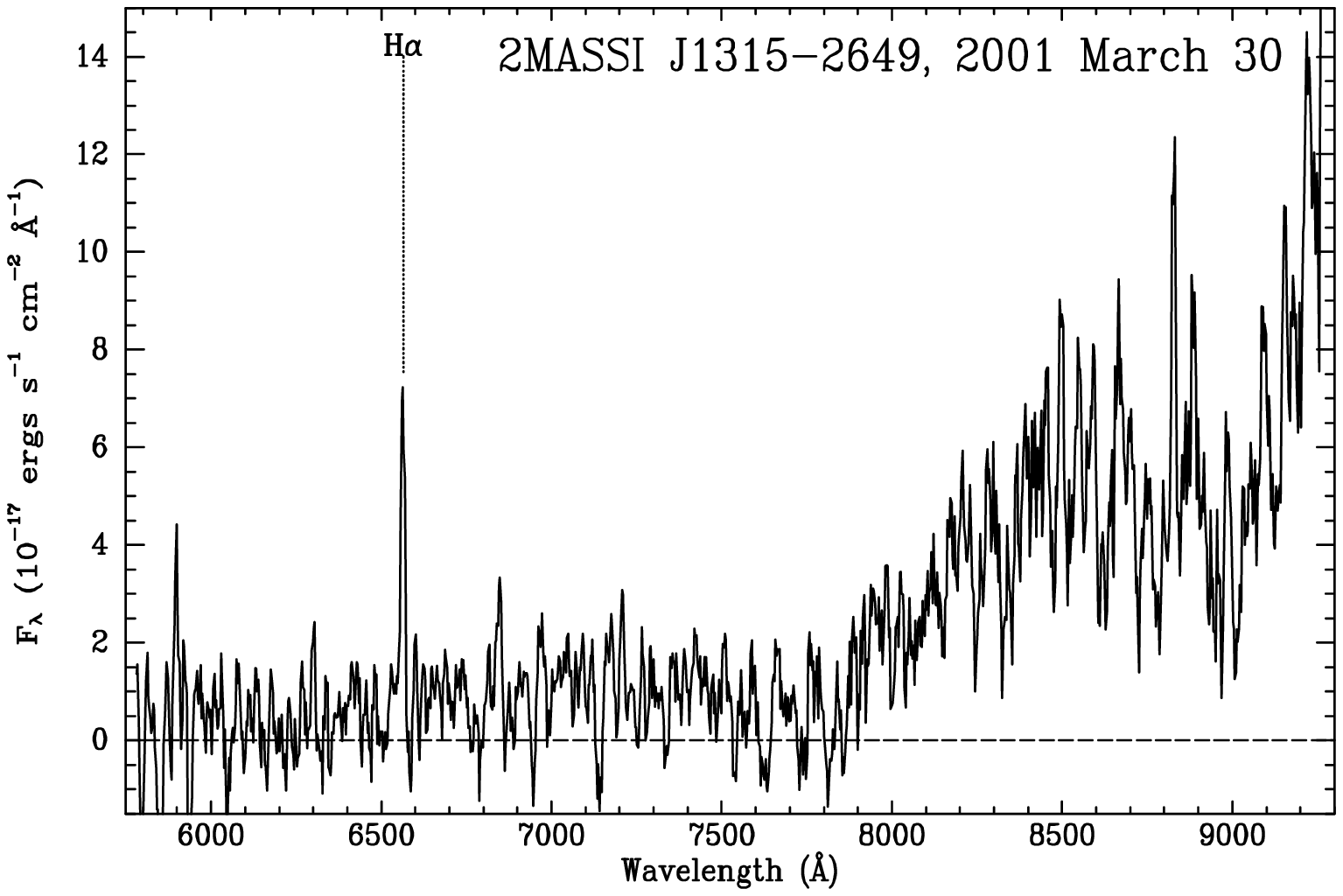}{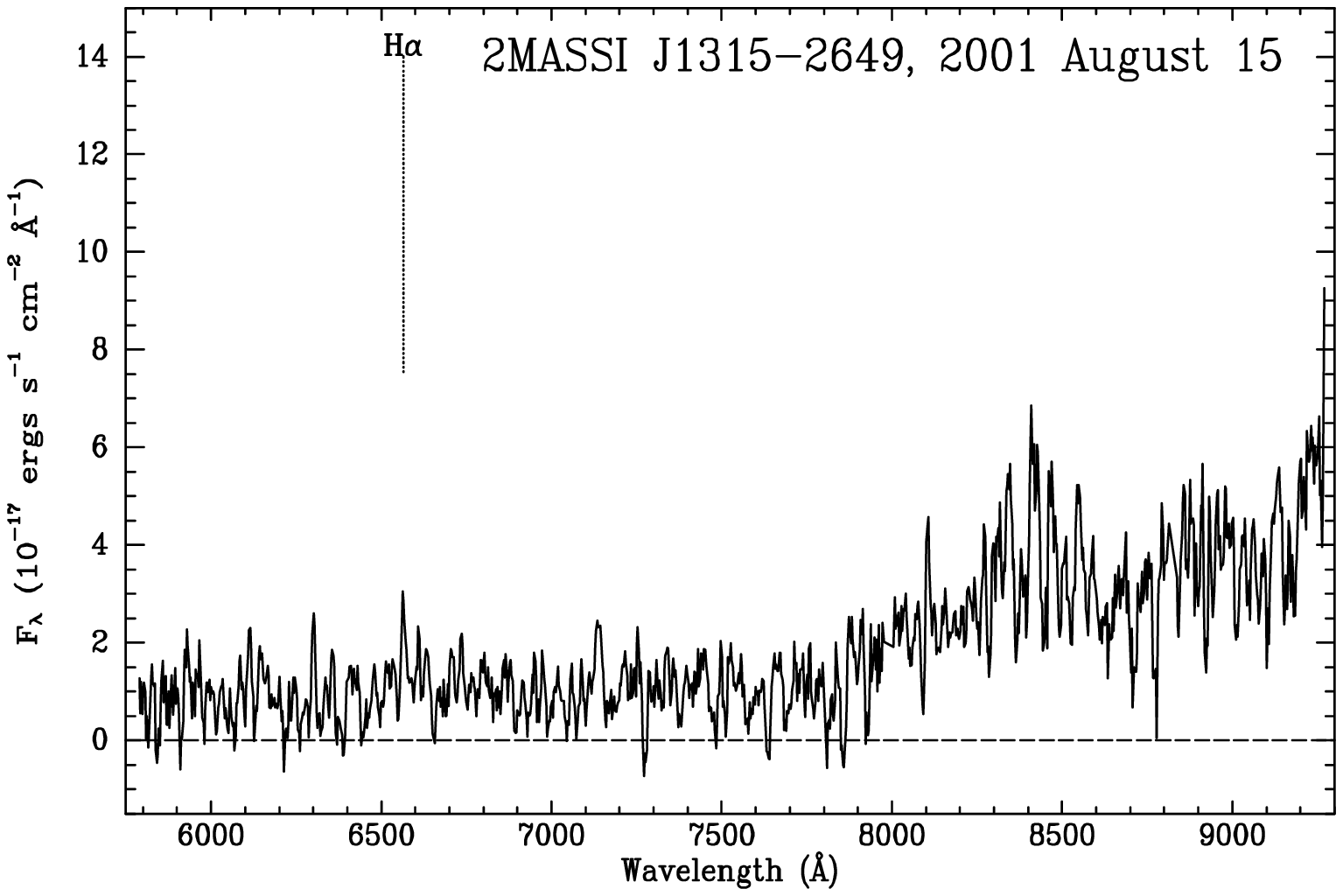}
\caption[]{ 
Optical spectra of 2MASSI J1315309$-$264951
obtained in (a) March and (b) August 2001.
Both spectra have been smoothed by a five pixel boxcar, 
and the dashed lines show the zero level for each spectrum.
The August spectrum was obtained in nonphotometric conditions, but nonetheless
it is clear that the \ha\ emission
was greatly diminished compared to March.
}\label{figure}
\end{figure}

Figure \ref{figure} shows the March and August spectra of 2MASS\,1315$-$2649.
By comparison to the spectral atlas of \markcite{kir00}{Kirkpatrick} {et~al.} (2000), I estimate a spectral
class between L2 and L4 on the \markcite{kir99}{Kirkpatrick} {et~al.} (1999) system, and adopt L3. 
Due to the poor quality of the spectra, this classification is not based on
single features, but on the agreement between numerous features.
The absence of strong TiO and VO features at 7000-8400\,\AA\ puts the class
no earlier than L2.  The flatness of the spectrum from 8400-9000\,\AA\ and
the continuum level there compared to 7500\,\AA\ are
inconsistent with spectral types L5 and higher.

\ha\ is visible in both the two-dimensional and one-dimensional spectra from
both observing runs, but it is clear that 
at least its equivalent width (EW) 
varied between the two runs.  The total
H$\alpha$ flux may not have been measured accurately in either run because the
east-west slit was comparable in size to the seeing and was not centered
precisely on 2MASS\,1315$-$2649.
Also, the August spectrum was obtained through cirrus 
and so its absolute fluxing is untrustworthy.
On the other hand, EW 
measurements from both runs will be accurate unless the H$\alpha$ emission is 
extended, for which there is no evidence in the two-dimensional spectra, or the
contamination of the L dwarf spectrum by the USNO star spectrum was different.

To estimate the contamination of the L dwarf spectrum by the USNO star
spectrum, in each two-dimensional spectrum the L dwarf aperture was
reflected in the spatial direction around the centroid of the USNO star
aperture and the flux in the mirrored aperture was extracted.  
This contaminating flux 
is negligible at $\gtrsim$8000\,\AA\ in both epochs,
but is larger at $\lesssim$8000\,\AA\ in the August spectrum, yielding
the stronger continuum seen there in Figure \ref{figure}b.  
This makes the August EW 
measurement an underestimate, 
but does not affect the spectral typing of the object.

The H$\alpha$ EW was 
121$\pm$31\,\AA\ 
in March but only
25$\pm$10\,\AA\ 
in August.
The H$\alpha$ flux was measured to be
3.23$\pm$0.82 $\times$ 10$^{-16}$ ergs s$^{-1}$ cm$^{-2}$ in March.
Even including a correction factor of 1.42 for the nonphotometric conditions of
the August measurement estimated by scaling the flux from 8000-9260\,\AA\ to
match the March data, the August \ha\ flux is only
1.49$\pm$0.58 $\times$ 10$^{-16}$ ergs s$^{-1}$ cm$^{-2}$,
about a factor of two below the March measurement.
Thus it does appear that
the absolute flux as well as the equivalent width of the \ha\ emission varied.

%
The ratio of the \ha\ luminosity (or flux) to the bolometric luminosity (or
flux) is a useful quantity since it can constrain the emission mechanism.
The bolometric flux of 2MASS\,1315$-$2649 was estimated using the bolometric
correction from the $K_s$ band of 
BC$_K$=3.33 found by \markcite{tmr93}{Tinney} {et~al.} (1993) for the L4 dwarf GD165B.\footnote{The
bolometric correction is {\em not} always negative, as erroneously stated
on p. 381 of \markcite{aaq}{Cox} (2000), because the zero point of the magnitude system 
in which it is calculated 
can be different from that of the bolometric magnitude system.}
This yields an apparent bolometric magnitude of $m_{\rm bol}=16.79$.  
Since $f_{\rm bol}=2.48\times10^{-5}$ ergs s$^{-1}$ cm$^{-2}$ for
$m_{\rm bol}=0$ \markcite{aaq}({Cox} 2000), 
$f_{\rm bol}=4.77\times10^{-12}$ ergs s$^{-1}$ cm$^{-2}$ for 2MASS\,1315$-$2649.
Thus $\log (L_{\rm H\alpha}/L_{\rm bol}) \gtrsim -4.17$ in March,
and $\gtrsim -4.51$ in August (after accounting 
for 
nonphotometric conditions).
Both values are 
lower limits due to the use of a narrow 
slit with uncertain placement relative to the object.

\section{Discussion}  \label{DISC}

How does the \ha\ emission in 2MASS\,1315$-$2649 compare to that in other
late-type dwarf stars? 
\markcite{giz00}{Gizis} {et~al.} (2000) studied the \ha\ emission properties of a sample of nearby M and
L dwarfs and found that $\sim$20$\pm$10\% of L2-L4 dwarfs show \ha\ emission.
However, the strongest \ha\ emission object in their 
sample\footnote{2MASSW\,J0326137+295015,
whose spectrum can be seen in Figure 16 of \markcite{kir99}{Kirkpatrick} {et~al.} (1999).}
has only $\log (L_{\rm H\alpha}/L_{\rm bol}) \simeq -5$.
Subsequently, \markcite{bur00}{Burgasser} {et~al.} (2000) reported 
\ha\ emission with $\log (L_{\rm H\alpha}/L_{\rm bol}) \simeq -4.3$ in the 
T dwarf 2MASSI\,J1237392+652615.
Possible variability in this object's \ha\ flux was reported at the 2.8$\sigma$
level in one spectrum, 
but more detailed analysis shows that the spectra are consistent with no
variability \markcite{bur02}({Burgasser} {et~al.} 2002).


Thus even the August spectrum of 2MASS\,1315$-$2649 has stronger
\ha\ emission than reported in any other L dwarf, and the
variable \ha\ emission in this object is the first reported for any L dwarf.
The study of \markcite{giz00}{Gizis} {et~al.} (2000) indicates that as an L dwarf with \ha\ emission,
2MASS\,1315$-$2649 is likely 
massive enough (and old enough) to have burned lithium,
but the physical explanation for this correlation with mass is unknown.
The various explanations considered by \markcite{bur00}{Burgasser} {et~al.} (2000) for \ha\ emission in 
the T dwarf 2MASS\,1237+6526 
can also be considered for 2MASS\,1315$-$2649.
Acoustic heating is ruled out as a possible dominant energy
source for strong \ha\ emission in L or T dwarfs \markcite{bur00}(\S3.3 of {Burgasser} {et~al.} 2000),
but several other possibilities remain viable.

\subsection{Accretion In An Interacting Binary System}

\markcite{bur00}{Burgasser} {et~al.} (2000) discuss the possibility of sustained Roche lobe overflow in close
brown dwarf binaries.  Such accretion might explain 
steady \ha\ emission without a
strong accompanying thermal spectrum if the accretion produces \ha\ emission
from ionization at a shock front or from magnetic field lines streaming onto the
pole of the primary, rather than from an accretion disk.  \markcite{bur00}{Burgasser} {et~al.} (2000) suggest
that the T dwarf 2MASS\,1237+6526 and the M9.5e dwarf PC\,0025+0447 could be
examples of such systems.  
Such an explanation cannot be ruled out for 2MASS\,1315$-$2649, but
its confirmed variable \ha\ 
would require variability in some part of the accretion process.
Such variability has not been convincingly observed in the other two objects
and might not be expected in the case of accretion from sustained
Roche lobe overflow.

\subsection{A Strong Magnetic Field}

Magnetic fields are believed to drive \ha\ emission from F to early M stars
via an internal dynamo which is stronger for faster-rotating stars, such as
the $\alpha$-$\Omega$ dynamo \markcite{par55}({Parker} 1955).  
Such dynamos require a radiative/convective boundary to 
anchor flux lines and therefore break down as stars become fully convective 
($\sim$0.3\,$M_{\odot}$, spectral type M4).
However, the observed \ha\ activity level in M dwarfs shows no sign of this
transition, remaining constant at $\log (L_{\rm H\alpha}/L_{\rm bol})\simeq-3.8$
\markcite{hgr96}({Hawley}, {Gizis}, \& {Reid} 1996).  This suggests that some other mechanism,
probably a turbulent dynamo \markcite{ddr93}({Durney}, {De Young}, \& {Roxburgh} 1993), contributes substantially
to the magnetic field in very late type dwarfs.

The value of $\log (L_{\rm H\alpha}/L_{\rm bol})$ in 2MASS\,1315$-$2649 is an
order of magnitude larger than in most L dwarfs, but is still 
lower than the average value of $-$3.8 measured for M dwarfs (see Figure 7 of
\markcite{giz00}{Gizis} {et~al.} 2000).
%
Thus it is at least plausible that the same mechanism at work in late M dwarfs
(whether a turbulent dynamo or something else)
could provide the magnetic field required to explain
the \ha\ flux in 2MASS\,1315$-$2649 and possibly the T dwarf 2MASS\,1237+6526.
This would mean that they just have stronger than average magnetic fields
for L and T dwarfs.  

If this is the case, then the variable \ha\ emission in 2MASS\,1315$-$2649
requires at least a slow variation in its magnetic field.
However,
no late type dwarf is known to show long-term variability of more than a factor
of two at X-ray or extreme ultraviolet wavelengths (\S\,4 of \markcite{dra96}{Drake} {et~al.} 1996),
and variability in \ha\ is not likely to be greater than variability at those
wavelengths.
On the other hand, flares are known to occur in M dwarfs and 
have been detected at radio wavelengths in
the L3.5 dwarf 2MASSW\,J0036159+182110 \markcite{ber02}({Berger} 2002).
Thus while the spectra presented here do not rule out a slow variation in its
magnetic field as the cause of the \ha\ variability in this L dwarf, a more
rapid variation --- a flare --- seems much more likely, as discussed next.
%
%
%

\subsection{Flaring}

Flares (seen in \ha, the radio, the UV, and X-rays) on late type stars are 
powered by the energy released
during a sudden reconfiguration of the magnetic field structure.  Detection
of a flare therefore confirms the existence of a stellar magnetic field.
However, strong flares in M dwarfs typically last only a few hours \markcite{hp91}({Hawley} \& {Pettersen} 1991) 
and occur unpredictably, making them difficult to observe.

It may very well be that 2MASS\,1315$-$2649 was flaring when the discovery
spectra were obtained.  
The \ha\ emission did not vary more than 1$\sigma$ (25\%) on a timescale
of $\sim$15 minutes between successive two-dimensional spectra in March,
but this is consistent with the rate of change of the \ha\ flux 
in the M9.5 dwarf 2MASS\,0149+2956 during the flare observed by \markcite{lie99}{Liebert} {et~al.} (1999).
%
Given that \ha\ flaring is known to occur in M dwarfs and radio flaring in 
at least one L dwarf \markcite{ber02}({Berger} 2002), flaring seems the most likely
explanation for the \ha\ variability in 2MASS\,1315$-$2649.
However, the other possibilities discussed earlier cannot yet be ruled out.
The current data does not prove that the variation in \ha\ occurred
on a timescale short enough to be called a flare,
and radio, \ha\ and X-ray and activity may be decoupled
for spectral types later than M8 \markcite{ber02}({Berger} 2002),
Further monitoring of 2MASS\,1315$-$2649 is needed to firmly 
determine the energy source for its \ha\ emission.


If its variable \ha\ emission is 
due to flaring, 2MASS\,1315$-$2649
is probably similar to the M9.5 dwarf BRI\,0021$-$0214 \markcite{rei99}({Reid} {et~al.} 1999).
Both objects show only \ha\ in emission and have 
peak $\log (L_{\rm H\alpha}/L_{\rm bol})$ values 
lower than the 
average for M dwarfs.
However, the spectra do not 
rule out the possibility that 2MASS\,1315$-$2649 is 
a flare star with 
emission from lines other than \ha, similar to 2MASS\,0149+2956 \markcite{lie99}({Liebert} {et~al.} 1999).
There is no evidence for such lines in March to limits of 
0.5$f_{{\rm H}\alpha}$, 
but in 2MASS\,0149+2956 the strongest other lines had 
0.1$f_{{\rm H}\alpha}$. 
There is also no evidence in the March spectra for the stronger continuum and
veiling of molecular bands seen during the flare in 2MASS\,0149+2956.  
Only changes of a factor of two or more in the continuum level can be ruled out,
but this is sufficient to place 2MASS\,1315$-$2649 at either an
earlier or later stage in any such flare, as follows.
The \ha\ EW in 2MASS\,0149+2956 increased (from 200\,\AA\ to 330\,\AA)
during the observed portion of the flare because the continuum flux at
\ha\ decreased more quickly (from ten to three times the quiescent flux)
than the \ha\ line flux did.
In 2MASS\,1315$-$2649 the \ha\ EW was 121\,\AA\ in March, but the continuum
flux at \ha\ was at most twice 
the quiescent flux (estimated from the August spectrum).
This \ha\ EW and continuum level cannot be simultaneously reproduced by simply
scaling any of the 2MASS\,0149+2956 observations, even accounting for the
weaker continuum at \ha\ in an L3 dwarf.
Thus, if they were the same type of flare, and if the relationship between the
continuum and \ha\ line flux in this type of flare is independent of its
luminosity, then 2MASS\,1315$-$2649 must have been observed at a later
(or perhaps earlier) stage of its flare.  
Any model for such flares must then incorporate the observed limit of $\sim$25\%
variation in \ha\ on $\sim$15 minute timescales at such a stage in the flare.

If the \ha\ emisssion in 2MASS\,1315$-$2649 in March was from a flare,
regardless of what type, 
then the large range in $\log (L_{\rm H\alpha}/L_{\rm bol})$ for flaring 
M dwarfs 
extends to L dwarfs as well.  
As pointed out by \markcite{rei99}{Reid} {et~al.} (1999), this suggests 
a wide variation in intrinsic magnetic field strength, efficiency of energy
transport to the stellar chromosphere during a flare, or both.

\section{Conclusion}  \label{CONCL}

2MASSI\,J1315309$-$264951 is an L3 dwarf with strong H$\alpha$ emission 
which decreased in strength by about a factor of two 
between two epochs separated by 137 days,
the first reported variable \ha\ emission in an L dwarf.
The \ha\ emission in 2MASSI J1315309$-$264951
must be powered either by magnetic activity or by accretion in a binary system.
Accreting binaries are rare, so that hypothesis is unlikely.
The spectra presented here do not rule out a slow variation in \ha\ strength,
but slow variations of the observed amplitude are rare among M dwarfs.
Since flaring powered by reconnection of magnetic fields is common in M dwarfs
and a radio flare has been detected in the L3.5 dwarf 2MASSW\,J0036159+182110
\markcite{ber02}({Berger} 2002), a flare is the logical explanation for the \ha\ variability in
2MASSI\,J1315309$-$264951, 

The value of $\log(L_{\rm H\alpha}/L_{\rm bol}) > -4.17$
observed at the discovery epoch is larger than that of any other L dwarf 
but comparable to the value of $-4.3$ observed for
2MASSI J1237392+652615, the only reported T dwarf with \ha\ emission.
However, both these values lie well below the average
$\log(L_{\rm H\alpha}/L_{\rm bol})=-3.8$ observed in M dwarfs.

Only two L dwarfs and one T dwarf are known to exhibit \ha\ emission of
strength $\log(L_{\rm H\alpha}/L_{\rm bol}) > -5$.  Thus perhaps two percent of
L or T dwarfs exhibit \ha\ emission this strong at any given time.
Given the small number statistics, this is consistent with the duty cycles
observed for \ha\ flares ($\sim$7\%; \markcite{giz00}{Gizis} {et~al.} 2000)
and radio flares (2-10\%; \markcite{ber02}{Berger} 2002) among late-type dwarfs.
Spectroscopic or narrow-band \ha\ monitoring of L and T dwarfs on timescales
of hours to days is needed to determine if the frequency of strong \ha\ emission
is governed by flaring, by the upper envelope of the magnetic field strength
distribution,
or by accretion in binary systems.

\acknowledgements

I thank an anonymous referee and A. Bugasser for helpful comments,
CNTAC for observing time, A. Alvarez and A. Guerra for operating the CTIO 1.5m
and putting up with my musical tastes, 
CTIO for excellent observing support, and Fundaci\'{o}n Andes and Chilean 
FONDECYT grant \#1010981 for financial support.
The Two Micron All Sky Survey (2MASS) is a joint project of the University of
Massachusetts and the Infrared Processing and Analysis Center/California
Institute of Technology, funded by the National Aeronautics and Space
Administration and the National Science Foundation.


\begin{deluxetable}{cccccccccc}
\tablecaption{Properties of 2MASSI J1315309$-$264951\label{t_info}}
\tabletypesize{\scriptsize}
\tablewidth{475.00000pt}
\tablehead{ 
\colhead{} & \colhead{} & \colhead{} & \colhead{} & \colhead{} & \colhead{} &
\colhead{} & \colhead{} & \multicolumn{2}{c}{H$\alpha$ EW, \AA}\\[.2ex]
\colhead{$B$} & \colhead{$R$} & \colhead{$J$} & \colhead{$H$} & 
\colhead{$K_s$} & \colhead{$J-H$} & \colhead{$H-K_s$} & \colhead{$J-K_s$} & 
{March 30} & {August 15} 
}
\startdata
$>$18.90 & $>$17.70 & 15.18$\pm$0.05 & 14.06$\pm$0.04 &
13.46$\pm$0.04 & 1.13$\pm$0.06 & 0.60$\pm$0.05 & 1.73$\pm$0.06 
& 121$\pm$31 	
&  25$\pm$10 \\ 
%
\enddata
\tablecomments{The 2MASS database is matched to the USNO-A2.0 catalog within a
radius of 5\farcs0, and so the optical magnitudes of USNO\,J131531.23$-$264953.0
are listed in the 2MASS database entry for 2MASS\,1315$-$2649 
(see \S\ref{SPEC}).  We list them here merely as lower limits
to the optical magnitudes of 2MASS\,1315$-$2649.
}
\end{deluxetable}

\end{document}